# A New Approach for the Fabrication of MgB$_2$ Superconducting Tape with Large In-field Transport Critical Current Density

K. Komori, K. Kawagishi, Y. Takano, S. Arisawa, H. Kumakura and M. Fukutomi
*National Institute for Materials Science, Superconducting Materials Center, 1-2-1, Sengen, Tsukuba, Ibaraki 305-0047, Japan*

K. Togano
*National Institute for Materials Science, Fundamental Materials Research Center, 1-2-1, Sengen, Tsukuba, Ibaraki 305-0047, Japan and CREST, Japan Science and Technology Corporation, 2-1-6, Sengen, Tsukuba 305-0047, Japan*

The paper reports the first successful fabrication of MgB$_2$ superconducting tape using a flexible metallic substrate as well as its strong pinning force, which was verified by direct measurement of transport critical current density. The tape was prepared by depositing MgB$_2$ film on a Hastelloy tape buffered with an YSZ layer. The $J_c$ of the tape exceeds $10^5$ A/cm$^2$ at 4.2K and 10T, which is considered as a common benchmark for magnet application. The $J_c$ dependence on magnetic field remains surprisingly very small up to 10T, suggesting that the tape has much better magnetic field characteristic than conventional Nb-Ti wires in liquid helium.

The newly discovered MgB$_2$ superconductor[1] is expected to be useful for various electric power applications as well as electronic device applications because its transition temperature is much higher than those of conventional metallic superconductors such as Nb-Ti and Nb$_3$Sn. In order to evaluate the potentiality for power applications, the development of wire processing techniques is essential. The first attempt at wire fabrication was the magnesium vapor diffusion to boron fibers presented by Canfield et al[2]. Recent efforts at wire fabrication have centered on developing the powder-in-tube (PIT) process due to its greater easiness of scaled up production[3-7]. The transport critical current density $J_c$ at 4.2K and in self-field of the PIT processed tapes has already exceeded the practical level of $10^5$ A/cm$^2$. However, the $J_c$ rapidly decreases in an applied magnetic field due to its weak pinning force. The $J_c$ values at 4.2K and 10T reported so far for PIT processed wires and tapes remain as low as the order of $10^3$ A/cm$^2$, although the values are being steadily improved.

In contrast to the relatively poor in-field $J_c$ values of the tapes and bulks, there are several papers on MgB$_2$ thin films, which reported very much higher Hc$_2$ values and $J_c$ values in applied magnetic fields than those of tape and bulk samples[8,9]. Kim et al[9] reported a large transport $J_c$ value of $10^5$ A/cm$^2$ at 5T and 15K for the c-axis oriented MgB$_2$ film prepared on an Al$_2$O$_3$ substrate. Those results suggest to us that the MgB$_2$ phase formed by vapor deposition techniques has extremely strong pinning force. However, all experiments on the thin films have been performed using ceramic substrates, which are not suitable for long length production of flexible conductor.

This paper reports a successful new approach to fabricate MgB$_2$ tape that is the combination of the high critical current density achieved by vapor deposition and the use of flexible metallic substrate tape. The technique is similar to the so-called coated conductor techniques developed for YBa$_2$Cu$_3$O$_y$(Y-123), where Y-123 thin films are deposited on various buffer layers on a metallic substrate[10-12]. The MgB$_2$ tape obtained in this work has an excellent transport $J_c$ of $1.1 \times 10^5$ A/cm$^2$ at 4.2.K and 10T. It is also surprising that the $J_c$ dependence on the magnetic field remains extremely small up to 10T, indicating that the tape has a much better high magnetic field characteristic than conventional Nb-Ti wires. This is the first demonstration of a high critical current capability of MgB$_2$ tape by actual transport measurements in strong magnetic fields.

The samples were prepared as follows. The substrate of Hastelloy(C-276) tape, 4x30x0.3 mm$^3$, was mirror-like polished and pre-coated with an about 1μm thick YSZ(yttria-stabilized zirconia) buffer layer by bias sputtering[13]. The buffered-substrate tapes were mounted on a heater plate using silver paint. For deposition of MgB$_2$ films, a KrF excimer laser (λ=248nm) with 400 mJ/pulse operating at 5 Hz was used. In order to compensate the losses in Mg during post-deposition anneals, a magnesium rich target was prepared by mixing MgB$_2$ powder ( Alfa Aesar ) and Mg powder (purity 99.9%) and pressed into a pellet. The ratio of Mg to B was set to 1.5~2.5 : 1 molar ratio. The films were deposited on unheated substrates in a 6x10$^{-5}$ Torr Ar atmosphere for 40 min. The films were then heated for 20~30 min in the temperature range of 550 to 660°C followed by a quick cooling to room temperature. In this annealing step, the color of the film changed from bright silver to shiny black. In addition, the film thickness also changed from about 800 nm to about 400nm as already observed by Zeng et al[14].

In the x-ray diffraction pattern, no discernible peak corresponding to MgB$_2$ phase was seen except for the peaks of YSZ and Hastelloy, suggesting that the MgB$_2$ phase has a very small grain size[14, 15]. **Figure 1** shows a typical cross section of the film exposed by fracturing the sample and observed with a scanning electron microscope (SEM). The film is apparently composed of two layers of YSZ buffer and MgB$_2$. The average thickness of the MgB$_2$ layer is around 0.4μm in accordance with the result of thickness measurement by a stylus profilometer. The MgB$_2$ film is very adherent to and uniformly formed on the YSZ buffer layer. Succeeding to the columnar structure of the YSZ buffer layer, the MgB$_2$ layer also appears to grow with somewhat columnar shape whose size is about 50nm. Microstructure observation by a transmission electron microscope (TEM) was also performed using JEOL-JEM-4000EX. The TEM specimen was prepared by crushing the flakes of deposited layer scraped off from the Hastelloy substrate. **Figure 2** shows the bright field image of MgB$_2$ layer and corresponding electron diffraction pattern, which indicate that the structure is composed of the MgB$_2$ matrix and the particles of MgO phase with apparent lattice image whose size is around 10nm. Careful observation showed that the MgB$_2$ matrix is composed of very fine grains whose size is less than 10nm, although the grain boundary is not clear in Fig.2.

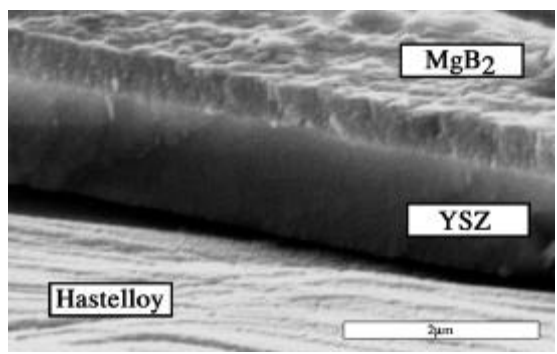

Figure 1 Scanning electron micrograph observed on the fractured cross section of the MgB$_2$ tape.

**Figure 3** shows a typical temperature dependence of the resistance for two MgB$_2$ tapes, tapes 1 and 2, which were prepared under identical conditions except for the target composition, Mg/B=1.5 (tape 1) and 1.2 (tape 2). Both tapes show a sharp superconducting transition with the zero resistance transition temperatures of 29 K and 28K for tapes 1 and 2, respectively. These values are much lower than the ideal value of 39K reported for bulk sample, in accordance with the past reports of post annealed films grown from sintered targets[16-19]. The resistance ratio defined by $RR=R(290K)/R(T_c)$ varied between the two films; 1.0 for tape 1 and 0.6 for tape 2. These small $RR$ values, as is often observed in the film prepared with an in situ two-step process, may be caused by very fine structure as observed in Fig.2. The electrical insulation by the YSZ buffer layer is not complete probably due to the presence of pinholes in the YSZ layer, which is responsible for relatively small resistance of mΩ order compared to that expected from only the resistance of MgB$_2$ film.

In order to determine magnetic parameters such as upper critical field, $H_{c2}$, and irreversibility field, $H_{irr}$, the resisitive superconducting transitions were measured under magnetic fields up to 9T applied parallel to the tape surface. **Figure 4** shows the plot of $H_{c2}$ and $H_{irr}$ for tape 1 as a function of temperature, which were determined as the onset and offset temperatures of the transition, respectively. Linear extrapolation gives the $H_{c2}$ (4.2K) and $H_{irr}$ (4.2K) values of about 33T and 18T, respectively. These are very high values exceeding even those of single crystals in the magnetic field parallel to the a-b plane[20].

Measurements of the field-dependent critical current $I_c$ ($H$) were performed in liquid helium and magnetic fields up to 12T generated by a superconducting magnet. The applied field direction was parallel to the tape surface. In order to reduce the measuring current, we fabricated in situ patterned MgB$_2$ films on YSZ-buffered Hastelloy tape by shadow masking during PLD. The pattered film thus obtained had a stripe of about 1.2mm width and 5mm length in the middle of the tape. Voltage and current leads were attached to the sample using In-Ga alloy. The voltage taps were attached to both the MgB$_2$ surface and the Hastelloy surface on the reverse side in order to avoid the overestimation of $I_c$ due to current sharing with the Hastelloy tape. **Figure 5** shows typical voltage versus current curves of tapes 1 and 2 measured by fixing the applied magnetic field and sweeping the current. There was no voltage appearance within the detection limit up to $I_c$, indicating that no appreciable current sharing occurred while the MgB$_2$ layer kept zero-resistance state. The catastrophic voltage appearance observed for tape 2 when $I_c$ exceeded a certain value is probably due to the heating of the contact and hence those were omitted from the $J_c$ calculation. $I_c$ were defined by a potential drop of 1μV appeared across the voltage probes of 8mm distance. The $J_c$ was calculated by dividing the measured $I_c$ value by the cross sectional area of the MgB$_2$ layer.

**Figure 6** shows the $J_c$-$H$ curves for tapes 1 and 2. For comparison, typical curves of PIT processed $MgB_2$ tapes[6,7] and a Nb-Ti superconductor[21] are included in the figure. It is noticed that tapes 1 and 2 show an extremely small magnetic field dependence of $J_c$ compared to those of the PIT processed tapes. Therefore, both tapes maintain very large $J_c$ values of more than $10^4 A/cm^2$ up to 12T. The highest $J_c$ value at 10T and 4.2K is $1.1 \times 10^5 A/cm^2$, which is about two orders of magnitude larger than the typical value of PIT processed tape. A possible mechanism of such small magnetic field dependence and high $J_c$ value in an applied magnetic field is a strong grain boundary pinning force. This can be speculated by analogy with the case of A-15 intermetallic compound superconductors for which strong correlations between $J_c$ and grain size were reported; the pinning force is inversely proportional to the grain size[22]. As mentioned earlier, the $MgB_2$ phase has an extremely fine grain size of less than 10nm. This is much smaller than the typical $MgB_2$ grain size of PIT processed tape, which is the order of 100nm-1000nm[6,7]. However, the possibility of pinning by oxygen and MgO incorporated in the film suggested by Eom et al[8] can not be excluded because the $MgB_2$ layer actually contains MgO particles as shown in Fig.2. Comparison of the results between tapes 1 and 2 indicated that a $J_c$ as high as the order of $10^5 A/cm^2$ (4.2K, 10T) can be attained for films with the $T_c$ and $RR$ values of about 30K and around 1.0. Systematic investigations on the relationship between well-defined microstructure and $J_c$ are underway in order to understand the pinning mechanism of the $MgB_2$ tape.

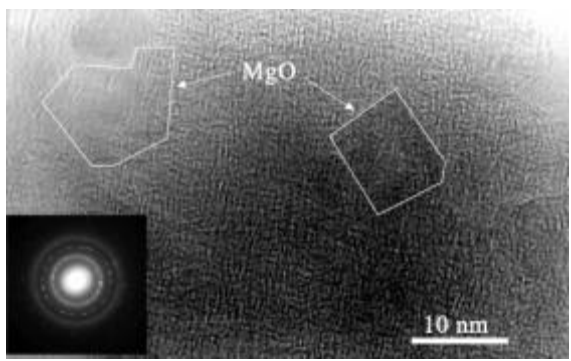

Figure 2  Transmission electron micrograph and corresponding electron diffraction pattern for MgB$_2$ layer of the tape.

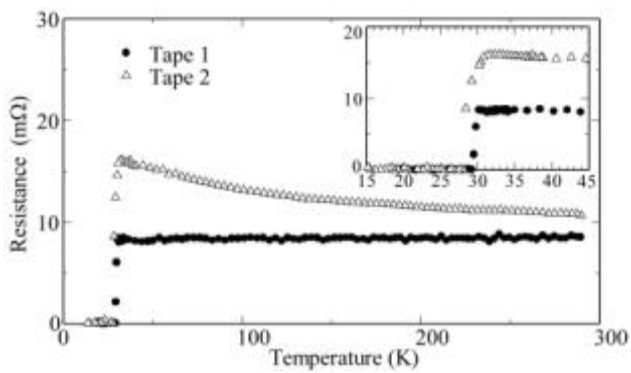

Figure 3  Resistance versus temperature curves of tape 1 and tape 2.

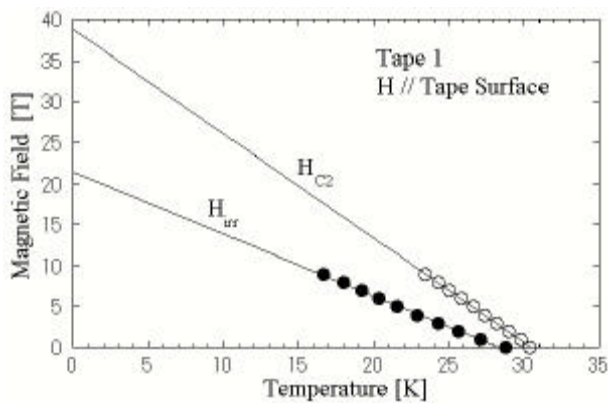

Figure 4  Temperature dependence of $H_{c2}$ and $H_{irr}$ determined by the onset and offset temperatures, respectively, of superconducting transitions in magnetic fields up to 9T for tape 1.

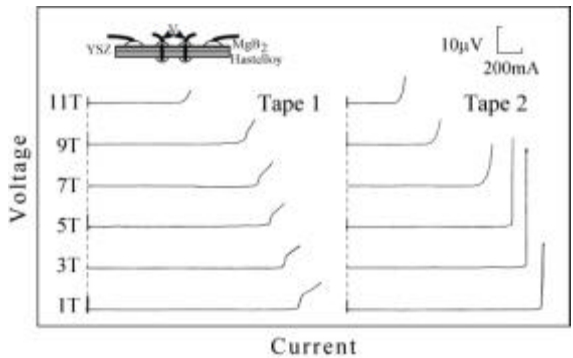

Figure 5 Typical voltage versus current curves for $I_c$ measurements of tape 1 and tape 2 measured by fixing the applied magnetic field and sweeping the current at 4.2K.

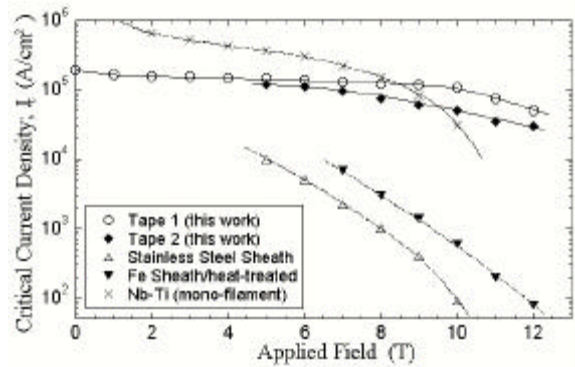

Figure 6 Transport $J_c$ at 4.2K of tape 1 and tape 2 as a function of applied magnetic field. Typical $J_c$-$H$ curves of PIT processed $MgB_2$ tapes[6,7] and a conventional Nb-Ti wire[21] are included for comparison.

In summary, we have demonstrated the first successful fabrication of $MgB_2$ thin film on a flexible metallic substrate as well as its strong pinning force. The transport $J_c$ at 4.2K and 10T exceeds $10^5 A/cm^2$, which is considered a common benchmark for magnet application. The $J_c$ dependence on magnetic field remains very small even at 10T, suggesting that the tape has much better magnetic field characteristic than conventional Nb-Ti wires in liquid helium. Of course, the scaling up of our technique is not so straightforward as the PIT process, but, the result shows that the $MgB_2$ superconductor has good potential for magnetic field applications.

The authors would like to thank to Dr. A. Matsumoto of National Institute for Materials Science for his helps in the $I_c$ measurement in applied magnetic fields.